\documentclass[journal]{IEEEtran}
\ifCLASSINFOpdf \else \fi

\usepackage{graphicx}

\usepackage[cmex10]{amsmath}
\usepackage{amssymb}

\usepackage{algorithmic}
\usepackage{array}
\ifCLASSOPTIONcompsoc
 \usepackage[caption=false,font=normalsize,labelfont=sf,textfont=sf]{subfig}
\else
 \usepackage[caption=false,font=footnotesize]{subfig}
\fi
\usepackage{url}
\usepackage[noadjust]{cite}

\begin{document}

\title{Architectural Design for Secure Smart Contract Development}

\author{\IEEEauthorblockN{Myles Lewis$^1$, Chris Crawford$^1$}
\IEEEauthorblockA{\textit{$^1$Dept. of Computer Science,} \\
\textit{The University of Alabama, Tuscaloosa, AL / USA} \\
mlewis16@crimson.ua.edu; crawford@cs.ua.edu}
}

\maketitle

\begin{abstract}
As time progresses, the need for more secure applications grows exponentially. The different types of sensitive information that is being transferred virtually has sparked a rise in systems that leverage blockchain. Different sectors are beginning to use this disruptive technology to evaluate the risks and benefits. Sectors like finance, medicine, higher education, and wireless communication have research regarding blockchain. Futhermore, the need for security standards in this area of research is pivotal. In recent past, several attacks on blockchain infrastructures have resulted in hundreds of millions dollars lost and sensitive information compromised. Some of these attacks include DAO attacks, bZx attacks, and Parity Multisignature Wallet Double Attacks which targeted vulnerabilities within smart contracts on the Ethereum network. These attacks exposed the weaknesses of current smart contract development practices which has led to the increase in distrust and adoption of systems that leverage blockchain for its functionality. In this paper, I identify common software vulnerabilities and attacks on blockchain infrastructures, thoroughly detail the smart contract development process and propose a model for ensuring a stronger security standard for future systems leveraging smart contracts. The purpose for proposing a model is to promote trust among end users in the system which is a foundational element for blockchain adoption in the future.
\end{abstract}


\IEEEpeerreviewmaketitle
\section{Introduction}

 Blockchain leverages decentralization techniques for database systems. The purpose of these systems is to promote trustless relationships between entities utilizing the systems, secure transmission and data storage, and the immutability of information stored onchain. Software development vulnerabilities in common blockchain infrastructures have been researched in \cite{chen2020survey, gupta2020smart, kushwaha2022systematic}.	Software vulnerabilities and threats within blockchain infrastructures typically can be located within the smart contract design phase \cite{wohrer2018smart}. New programming languages are being leveraged to automate the data transmission and state of information on chain by programmers. Solidity is a statically-typed programming language for the Ethereum blockchain. Pact is a programming language used for smart contract development on the Kadena network. Liquidity is a high-level typed smart contract language for the Tezos blockchain. Software development vulnerabilities can be located in the smart contract development stage since that is where programmers can develop code that can be implemented to handle data transmission and manipulation on-chain.

 \subsection{Types of Analysis}

 Smart contract security can be broken in two parts: logical errors and semantic misunderstandings from developers when programming. Also, Smart contract analysis can be broken in two parts: Static and Dynamic Analysis \cite{vivar2020analysis}. Static analysis pertains to the behavior of the program from its compiled binary code without executing the code. Observations can be made to look for patterns that can lead to vulnerabilities. 

 Dynamic analysis acts in the execution phase, detecting vulnerabilities that could have gone unnoticed during the static analysis \cite{vivar2020analysis}. By examining the security of the code in this stage of development through access control, encryptions, and vulnerabilities, we can further secure the data transmission and storage. The type of analysis relies on when the system is being evaluated in the development process. Most common forms of dynamic analysis found in literature pertains to execution tracing, symbolic analysis, validation of false/true positives, performance analysis, fuzz testing, and symbolic execution.
 
 Overall, these are just a few examples of the types of dynamic analysis that can be performed on smart contracts. Dynamic analysis can be an important tool for identifying potential vulnerabilities and improving the security and reliability of smart contracts. 

 \begin{figure}
     \centering
     \includegraphics[width=3.5in]{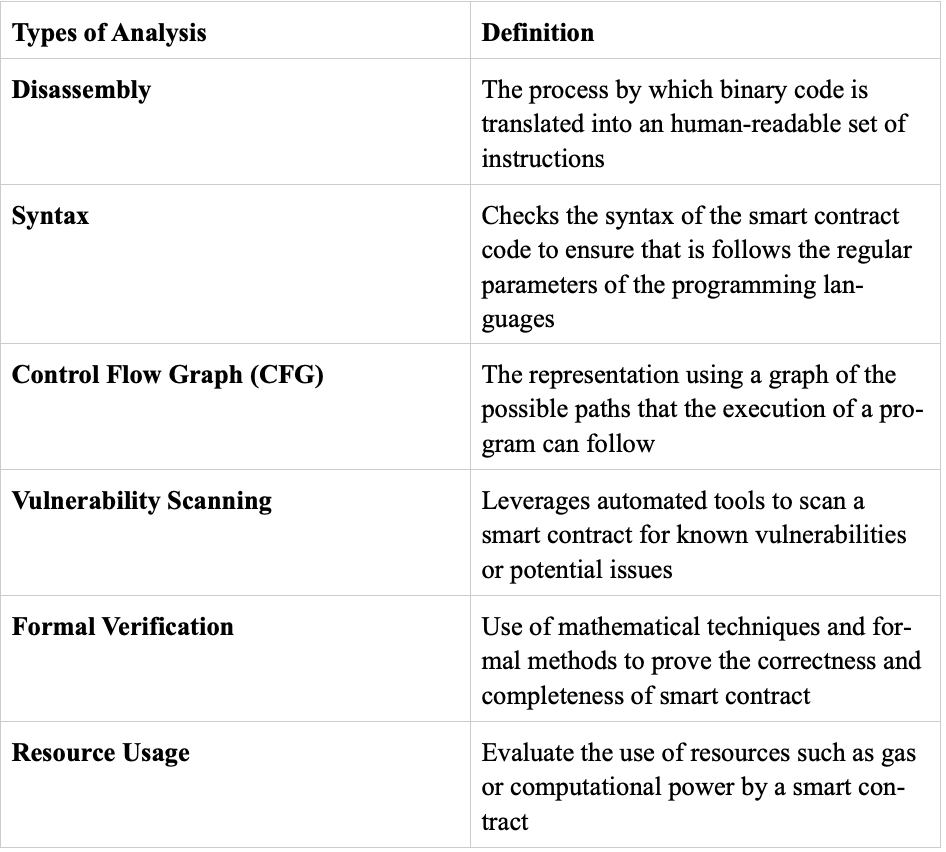}
     \caption{Static analysis types alongside their definitions}
     \label{fig:ToA}
 \end{figure}

\subsection{Types of Attacks}
Tools have been developed to find common software development vulnerabilities in smart contracts, Oyente \cite{luu2016making}. Oyente is used to find common vulnerabilities and notify the user for highly vulnerable smart contracts located onchain. The most common vulnerabilities that are found are: Callstack Depth Attack Vulnerability, Reentrancy Vulnerability, Assertion Failure, Timestamp Dependency, Parity Multigeniture Bug 2, Unchecked Inputs, Unhandled Exceptions and Transaction-Ordering Dependence. Unchecked inputs occur when smart contracts do not properly validate the input it receives from external sources outside of data stored on the network. Attacks can supply malicious input to cause the contract to behave outside of its initial functionality. Also, unhandled exceptions occurs when smart contracts do not properly handle exceptions or errors upon deployment. Attackers can then trigger unpredictable exceptions in the contract which can lead to crashes or unpredictable behavior. It is important for developers to be aware of these software vulnerability issues in blockchain frameworks by executing good security measure practices in their code, as much as needed.

\section{Related Works}
Blockchain technology has attracted significant attention in recent years, with many researchers exploring its potential applications and challenges \cite{chen2018exploring}. In this section, I review some of the key research in the field to provide valuable insight and guidance for future development in blockchain security protocols. 

\subsection{Smart Contract Security}
\cite{atzei2017survey} examined smart contracts on the Ethereum network to provide most common software programming is- sues that lead to vulnerabilities. The smart contracts evaluated were involved with current DeFi applications, to showcase the immediate need for a smart contract evaluation protocol to be in place. The most common issues is detecting mismatches between their intended behavior and the actual one, or vali- dating functionality. Verification tools can aid in this problem more quickly than developing a Turing-complete language. 
Oyente \cite{luu2016making}, can be used to detect vulnerabilities through static analysis of the contract code. More specifically, Oyente extracts the CFG through the Ethereum Virtual Machine (EVM) bytecode, then symbolically executed to find vulnerability patterns. Most common vulnerabilities found are exception disorder, time constraints, unpredictable state and reentrancy \cite{atzei2017survey}. Frameworks are being developed in the current state-of-the-art for application verification on blockchain infrastructures. \cite{bhargavan2016formal} proposed a functional programming language aimed at program verification, F*. F* is an example of static analysis, as it examines the contracts coded in solidity and EVM bytecode. By being given the solidity code and the correct compilation of the solidity code, the tool can verify that the two pieces of code have equivalent behaviors before being deployed onto the network.	
Other tools such as Zues, Osiris, Smartcheck \cite{tikhomirov2018smartcheck}, and Securify \cite{tsankov2018securify} exists to aid in strengthening software security in the smart contract development stage \cite{garfatta2021survey}. 

\subsection{Software Security Design}

Blockchain systems can utilize current smart contract security tools to create a conceptual security protocol design for future blockchain development. Security protocols are necessary in establishing a software security standard to ensure trust of nodes utilizing a system. This trust is essential in progressing blockchain technologies forward. But, there are limits in research that prohibit the terms of what makes a smart contract secure. For example, there are no concrete metrics for evaluating and comparing smart contracts, consistent smart contract code analysis or analytical coding methods that are easily digestible for traditional programmers \cite{vacca2021systematic}. 
An important component to constructing a secure smart contract is to carefully design and implement the contract using sound software engineering principles, and to thoroughly test the contract for potential vulnerabilities before deploying it to a blockchain network. This design should apply to all lifestyles of the smart contract, which include after deployment. 

\subsection{Blockchain Security}
Issues revolving around blockchain components such as consensus algorithms, cryptography, public ledgers can provide the necessary information that attackers need to attack the network. By using this information, attackers can perform various attacks that the smart contract will inherit because smart contracts are built on top of these blockchain networks. This is relevant to the work as the security of the underlying blockchain platform can contribute to the security of the smart contracts. 
There are several ways that security issues with a blockchain network can affect the smart contracts built on top of that network. 
First, if the blockchain network itself is not secure, it can make it easier for attackers to compromise the smart contracts running on that network. For example, if an attacker is able to gain control of majority of the nodes in the blockchain network, they can potentially disrupt the network’s operation, making it difficult or impossible for the smart contract to execute as intended. This is referred to as a 51\% attack in which hackers gain control of a majority of the network’s computing power and use it to manipulate the blockchain and execute unauthorized transactions \cite{aggarwal2021attacks}. 
Second, if the blockchain network is not transparent and immutable, it can make it more difficult for users to verify the security of the smarts running on that network. For instance, if a smart contract contains a bug, or vulnerability, it may not be immediately apparent to users, and this can create a risk of loss or damage for users of the contract. 
Third, if the blockchain network does not provide strong security guarantees for the execution of smart contracts, it can make it more difficult for users to trust the contracts running on that network. For example, if a smart contract is not executed in a secure, tamper-proof manner, users may be hesitant to use the contract, which can limit its usefulness and adoption. 
Ultimately, issues that can affect the security of blockchain can impact the security on smart contracts \cite{atzei2017survey}. It’s important for blockchain networks to prioritize security in order to ensure the integrity and trustworthiness of the smart contracts running on those networks. This can help to foster a more secure and stable environment for smart contract development and use. 

\section{Methodology}
In the literature, there is an extensive amount of resources into developing tools to analyze smart contracts, but there lacks a cohesive structure in utilizing these tools together. The structure can be used as a protocol that can validate and issue certificates for these evaluated smart contracts. The goal of my research is to design a conceptual model for a software security protocol for smart contract development. The key components for my approach involve combining static and dynamic analysis, identifying common software vulnerabilities, and distributing a security rating certificate. 

\subsection{Constructing Smart Contracts}
Typically from proper software programming practices, conceptualization is first. In constructing smart contracts, conceptualization is to define terms and conditions of the contract, functionality of the contract, requirements from the parties involved and expected outcome.
Next, the pseudocode for the smart contract can be coded. This is performed through a high-level programming language, such as Solidity, which is designed for constructing smart contracts. 
Following conceptualization and development of the contract, the code must be tested. The testing and debugging phase is to ensure that it is correct and free of errors. Practically, this debugging process is run on a local or test blockchain network that simulates various scenarios to verify that the contract behaves as expected. My approach begins to differ in this stage to apply static analysis on the smart contract alongside the debugging and testing on the testnet. 
The core steps in constructing a smart contract involve defining the contract’s terms and conditions, testing and debugging the contract’s code on the testnet, then applying static analysis. 

\subsection{Static Analysis}
Static Analysis is a process that is commonly used in soft- ware development to identify potential bugs, vulnerabilities, or other issues in the code without actually executing or deploying the code to interact with information on the main network of the blockchain. In the context of smart contracts, my approach to static analysis is to identify potential vulnerabilities in the contract’s code, such as improper use of cryptography, unhandled errors, issues that can compromise the security or integrity of the contract, and any of the other vulnerabilities previously mentioned in prior sections. 
There are several steps in my approach of performing static analysis on a smart contract. Initially, the contract’s code is analyzed using specialized tools and techniques to identify potential vulnerabilities or issues. These tools are specifically designed for smart contract development, such as Oyente or Zeus. Other general-purpose tools may exist that are used for static analysis in other contexts, but this research sticks to Ethereum based smart contract which require Ethereum based analytical tools. 
Subsequently, the results of the static analysis are to be reviewed and analyzed by a human, with considerable knowledge in blockchain security, to confirm the findings and determine the potential impact of any identified vulnerabilities. This can involve performing contract code reviews, testing the contract in simulated environments or test net cases, or consulting with other researchers in the field of smart contract and blockchain security to confirm the findings and assess the risks. 
Once the results of the static analysis have been reviewed and confirmed, a report is generated that summarizes the findings and provides two things: a recommendation of steps the developer can take to improve the security of the constructed smart contract and a evaluation report that details the security level of the contract under review. 
Overall, the process of static analysis for smart contracts development leverages specialized tools and techniques to identify potential vulnerabilities in the contract’s code, reviewing and confirming the findings, and generating a report with recommendations for addressing any identified issues. This stage of the process can help to improve the security and reliability of smart contracts and ensure they operate within the functionality that is intended for them.

\subsection{Security Rating Certificate}
The purpose of distributing a security rating certificate in the process of developing a smart contract is to ensure the security of the contract. This certificate is designed to provide evidence that the contract meets certain security requirements, such as being resistant to malicious attacks, and is properly configured to prevent unauthorized access. By distributing the certificate, it helps to ensure that the contract is secure and can be trusted by users and other entities interacting with it. By doing this, it helps to deter any potential malicious users from taking advantage of the contract.
In the context for smart contract development, the report has two purposes in my architectural design for this security protocol. First, provide the developer with potential bugs in the code that are recommended to be fixed prior to deployment. Then, provide the users that utilize the particular smart contract of its potential risks.
Since data is immutable on blockchain, this information is essential to gather before interacting with non-reversible, immutable data transmission on the main network. The report can serve as a performance evaluation, as it will alert the participating parties of what risks are evident in the smart contract. one of the key features of being blockchain, a security rating certificate can be distributed prior to deployment. Using the report that is generated during the static analysis phase, a rating can be placed upon the security level of the given smart contract. This rating can be combined with the smart con- tract so that users on the blockchain can view this rating prior to leveraging that smart contract.

\subsection{Deployment}
The deployment process for smart contracts involve the following steps: compiling the contract code, selecting a deployment network, creating and signing a deployment transaction, and confirming the contract deployment.
When compiling the code for Ethereum smart contracts, a Solidity compiler is used to convert the contract code written in Solidity’s programming language into bytecode that can be executed by the EVM. Once the code is compiled, a deployment network has to be chosen. In practical cases, developers would choose to directly deploy their code onto the main Ethereum network, but there are other options in place, a private or test network. It is encouraged that developers deploy to a test network at first to handle any unforeseen issues in the code.
Next, when deploying to a network, the developer will need to create a deployment transaction. Inside of the transaction, there will be the compiled contract code and any additional arguments or parameters. Developers will then need to sign and send the transaction. This is done by using the private key of the account that is deploying the contract. The signed portion of the transaction can then be compiled, by the EVM, with the code to ensure that the correct developer is uploading the smart contract.
Finally, the contract is deployed and available for use for any participating node on the network. By this process, developers will be required to run test on their smart contracts before they are eligible to deploy on to the mainnet. This will facilitate a reasonable standard as certain conditions will be required before public use of participating nodes.

\subsection{Dynamic Analysis}
Dynamic analysis is a process used to evaluate the behavior of a smart con- tract by executing it and observing its runtime behavior. This type of analysis can help identify potential vulnerabilities and security flaws in the contract code.
When identifying the inputs and outputs of the smart contract, it is important to understand the input-output behavior of the smart contract. This can be accomplished by comparing the types of data that the contract can accept as input and the types of values that the contract will output.
After conceptualizing the expected input-output behavior, test environments will need to be created when executing the contract. Again, in most blockchain networks, there is an implemented testnet that is ran to simulate the execution of these contracts. Practically, local blockchain emulators are essential but for Ethereum dynamic testing is done through the Rinkeby testnet.
Next, the developer should create test cases for the smart contract to gauge its functionality. Once the test environment is set up for the smart contract, the developer should include test cases that can cover a wide range of scenarios, involving normal and exception breaking samples.
Following the development of test cases, developers can run and observe the results on the smart contract and how the blockchain will theoretically react to these cases. This will allow the developer to understand the pitfalls of the execution of their smart contract with dynamic information interacting with the logical processes. Finally, the results can be analyzed to notice if there are any potential issues with the behavior of the contract.

\section{Future Works}
As with any complex software system, there is room for further development and improvement in the field of smart contract development. Some likely areas for future work: improved security, better debugging and testing tools, more efficient and scalable execution, and enhanced interoperability.
Improved security is one of the major challenges in smart contract development. There is a lack of transitional methods to ensure the security of contract code. Software security tools and practices are disjointed within the software development cycle for smart contracts. Static and Dynamic analysis, smart contract construction and deployment are not cohesive processes, but more separate, individual parts that are can be used by the developer.
Another area for future works is in the development of more advanced debugging and testing tools for smart contracts. These tools could provide better support for tracking the execution of the contract, identifying potential issues, and more comprehensible testing of the contract in variety of scenarios.
Also, as the popularity of blockchains grows, systems that leverage this particular component of blockchain will increase. This emphasizes the need for more efficient and scalable ways to execute these contracts on the main network of the blockchain. Future work can focus on developing new techniques and technologies that analyze smart contracts in dynamic analysis for prospective developers to improve the performance and scalability of the contract post deployment.
Enhanced interoperability refers to the ability for smart con- tracts to affect data on separate blockchain networks. This is a more advanced and futuri- stic goal as protocols and standards will definitely need to be in place for communication, data sharing and creation of tools for integrating contracts from different platforms.
In regards to the proposed security protocol, it is designed to address the major gaps in current security evaluation techniques for smart contract development. It provides a comprehensive and automated security audit to ensure the security of the contract and a blockchain-based security platform to protect against malicious actors. Additionally, the protocol can be used to create more efficient and cost-effective security evaluation techniques for smart con- tract development. By using the protocol, users can be confident that their contracts are secure and can be trusted by other entities interacting with it by having a structured process of smart contract security evaluations.
To improve my design for future implementation, I could explore new tools and techniques for ensuring the security of smart contracts. As I am exploring these alternative tools, I can compare, quantitatively, the results from each tool to find the most optimal route for smart contract development. Along- side comparing security tools for smart contracts, blockchain platforms will have to be evaluated themselves at some point. The security of the decentralized network plays a major role security of the smart contract or any other types of application built upon the network. The open source nature of blockchain platforms and the applications on these networks can cause the developers to be at a disadvantage when trying to secure their applications against hackers. Exploring blockchains that are not as public with the information as Ethereum will be important, but potential challenges can be raised as blockchain platforms are still relatively new and not every network has a designated programming language to develop applications. Also, explo- ring automated security audits and blockchain- based security platforms can be an extension of this work. Additionally, I could look into ways to make the protocol more efficient and cost-effective, such as using machine learning techniques to identify potential vulnerabilities quicker. Finally, I can investigate ways to make the protocol more user-friendly and accessible, such as by providing detailed documentation and tutorials.

\section{Conclusion}

The purpose of my paper is to contribute a conceptual security protocol for future development purposes for blockchain frameworks. By identifying common software vulnerabilities in the smart contract stage of the blockchain application development and leveraging evaluation tools found in research, a conceptual protocol can be constructed. The protocol can ensure a level of integrity for smart contract usage on the blockchain.
The problem within current literature is that each component discussed in the methodology section is modular and disjointed. I theorized a conceptual model of organizing the different components that are already in literature involving smart contract construction, static and dynamic analysis, smart contract deployment, and further monitoring of the contract while on the main network of the platform. My findings can provide the basis for constructing more complex and robust processes to collectively evaluate these contracts. Also, my work can lead to defining a standard for these smart contracts operating on these blockchain.

Additionally, researchers can develop new tools or techniques for detecting and mitigating security vulnerabilities. My solution was to create protocol that can fit the vast majority of current security tools and techniques to evaluate the majority of smart contracts. For some, this work can spark an idea to implement a more novel process with its own tools for static and dynamic analysis.
Finally, a future suggestion would be to investigate the potential for applying these principles to blockchain platforms outside of Ethereum. I am aware that different blockchains will leverage different static and dynamic analysis tools, but research possible ways to converge these methods into a singular process that can applicable for multiple blockchain platforms.

\begin{figure}
    \centering
    \includegraphics[width = 3.5in]{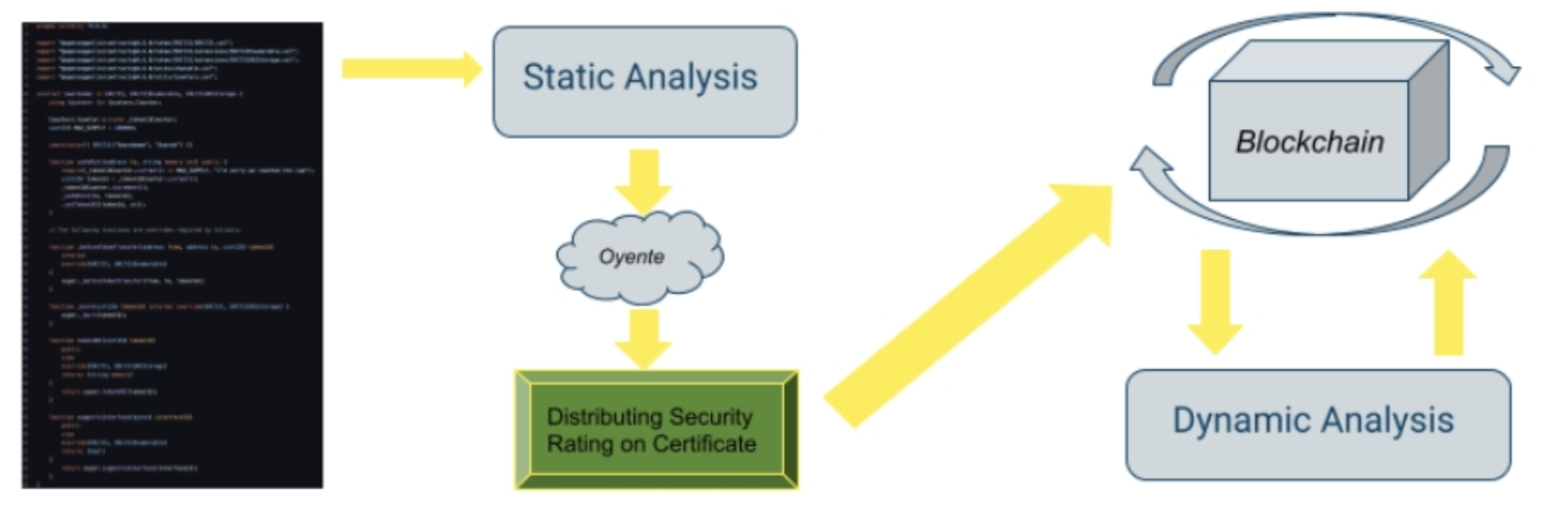}
    \caption{Architectural design for smart contract security model.}
    \label{fig:process}
\end{figure}

\bibliographystyle{plain}
\bibliography{bare_jrnl}
\end{document}